**Multicomponent rendezvous of cofilin, profilin and twinfilin at the actin filament barbed end**

Ankita[1], Sandeep Choubey[3,4,*] and Shashank Shekhar[1,2,*]

[1] Department of Physics, Emory University, Atlanta, GA 30322, USA

[2] Department of Cell Biology, Emory University, Atlanta, GA 30322, USA

[3] The Institute of Mathematical Sciences, Chennai 600113, India

[4] Homi Bhabha National Institute, Training School Complex, Anushaktinagar, Mumbai 400094, India

***Correspondence**: Shashank Shekhar (shekhar@emory.edu), Sandeep Choubey (sandeep@imsc.res.in)

**Abstract**

Cellular actin dynamics result from collective action of hundreds of regulatory proteins, majority of which target actin filaments at their barbed ends. Three key actin binding proteins - profilin, cofilin and twinfilin individually depolymerize filament barbed ends. Notwithstanding recent leaps in our understanding of their individual action, how they collectively regulate filament dynamics remains an open question. In absence of direct and simultaneous visualization of these proteins at barbed ends, gaining mechanistic insights has been challenging. We have here investigated multicomponent dynamics of profilin, cofilin and twinfilin using a hybrid approach that combines high throughput single filament experiments with theory. We discovered that while twinfilin competes with profilin, it promotes binding of cofilin to filament sides. Interestingly, contrary to previous expectations, we found that profilin and cofilin can simultaneously bind the same filament barbed end resulting in its accelerated depolymerization. Our study reveals that pair-wise interactions can effectively capture depolymerization dynamics in simultaneous presence of all three proteins. We thus believe that our approach of employing a theory-experiment dialog can potentially help decipher multicomponent regulation of actin dynamics.

**Main Text**

**Introduction**

Cells regulate assembly and remodeling of their actin cytoskeleton in response to mechanochemical signals (1-3). This response is mediated via the action of a large battery of proteins that combinatorially regulate dynamics of intracellular actin networks. Majority of these interactions take place at one of the two extremities of the actin filament, namely the barbed and the pointed end (1, 4). Decades of genetic experiments have identified key molecular components required for actin dynamics *in vivo*. Subsequent biochemical studies using purified proteins have since revealed how many of these components individually affect actin dynamics *in vitro*. Nonetheless, how cells integrate activities of these proteins to regulate actin assembly and disassembly of remains poorly understood.

There are three dominant approaches for deciphering multicomponent regulation of actin dynamics. Classically, the bulk pyrene fluorescence assay has been the technique of choice for investigating actin dynamics (5). While changes in pyrene signal readily allows measurement of effects of various actin binding proteins, this approach is difficult to employ for proteins which alter properties of the pyrene dye. For instance, cofilin binding to pyrene labelled F-actin causes quenching of the dye (6). The second approach involves direct visualization of fluorescently labelled protein molecules interacting with actin filaments using multicolor single-molecule microscopy. This technique has led to major discoveries such as mechanism of filament branching (7, 8), collaborative actin assembly by adenomatous polyposis coli (APC) and Spire (9), simultaneous association of elongators, blockers and depolymerases with actin filament barbed end (10-12), and synergistic depolymerization of pointed ends by cofilin and cyclase associated protein etc. (13, 14). Despite the great promise of this approach, labelling proteins without altering their activities and simultaneously visualizing multiple proteins is technically challenging and difficult to scale up. Further, this approach is difficult to employ for transient associations (<<1 s). The third approach is to leverage insights from structural studies of individual proteins bound to actin monomers (i.e., co-crystals) to predict how multiple proteins might simultaneously bind filaments and influence filament dynamics (10, 15, 16). However, since most of these co-crystal structures are for proteins bound to G-actin, majority of inferences are gained by docking structures of proteins bound to actin monomers on cryo-EM structures of actin filaments (10, 16). More recently, cryo-EM studies have allowed direct imaging of proteins bound to actin filament ends (17, 18).

Nevertheless, this approach has so far only been successful for individual proteins bound to filament ends with a very high affinity. As a result, these structure-based methods only provide a static picture taking into account only the long-lived binding configurations. Consequently, it has so far not been possible to employ this approach to investigate the effects of short-lived transient multiprotein complexes.

Here we present an alternative strategy of combining high-throughput experimental measurements with theory to study multicomponent regulation of actin dynamics without the need to directly visualize the individual proteins. Total internal reflection fluorescence microscopy (TIRF) has been the technique of choice for visualizing dynamics of individual actin filaments for over two decades (19, 20). More recently, combining microfluidics and TIRF has enabled rapid and precise measurement of actin filament end dynamics across hundreds of filaments (10, 13, 14, 21-29). Such high throughput quantitative measurements are amenable to theoretical modeling. To this end, falsifiable models predicated upon distinct mechanisms, which make specific predictions, can be built, and the corresponding predictions can be tested experimentally. To test the applicability of our theory-experiment approach, we asked how three key actin binding proteins - profilin, twinfilin and cofilin- together cause barbed-end depolymerization of actin filaments. Our results provide insights on whether these three proteins compete or cooperate at barbed ends to regulate filament depolymerization.

All three of these proteins can bind filamentous and monomeric actin. Their specific affinities and effects on barbed-end dynamics however are highly sensitive to the nucleotide state of actin (1, 3). Profilin preferentially binds ATP-G-actin subunits to inhibit filament nucleation, prevents monomer association at filament pointed ends (30-33), and supports processive elongation by delivering monomers to proline-rich proteins (formins, Ena/VASP (vasodilator-stimulated phosphoprotein) etc.) (34-36). Cofilin and twinfilin, which are both members of the ADF/Cofilin family of proteins, on the other hand bind aged-ADP-actin monomers much more strongly than ATP-actin monomers (6, 37-39). More recently, each of these proteins has been shown to bind filament barbed ends and accelerate depolymerization of newly-assembled ADP-$P_i$ actin filaments (16, 21, 23, 28). Nevertheless, how these three proteins influence barbed-end depolymerization when simultaneously present has never been addressed. We therefore applied our theory-experiment approach to investigate their combined interactions at ADP-$P_i$ filament barbed ends. We have here elected to investigate the effects of these proteins on ADP-Pi rather than ADP-actin barbed ends to avoid complications due to cofilin's rapid severing of ADP-actin filaments.

To gain mechanistic insights on multiprotein regulation of actin depolymerization by these three proteins, we first considered two broad classes of mechanisms via which any two simultaneously present depolymerases can interact with filament barbed ends. In the first mechanism, the two proteins compete at barbed ends, resulting in their mutually exclusive binding where only one of them is bound (and depolymerizing barbed ends) at a given time. In the second mechanism, the two proteins simultaneously bind to the same barbed end leading to increased rates of depolymerization. We developed an orthogonal approach by combining high throughput experiments with theoretical modelling to distinguish between these two modes of multiprotein interactions. Using multicolor single-molecule and high throughput microfluidics-assisted TIRF imaging (mf-TIRF) of hundreds of actin filaments, we first carried out a careful examination of barbed end depolymerization by these depolymerases either individually or in pairs. Results from earlier studies suggest that profilin and cofilin bind actin monomers in a mutually exclusive fashion due to their targeting of the same barbed surface of G-actin (40-42). However, we discovered that they are able to simultaneously associate with the same filament barbed end leading to enhanced depolymerization. Surprisingly, when actin filaments were simultaneously exposed to a solution containing cofilin and twinfilin, we found that observed depolymerization rates were consistently lower than those predicted by a model accounting for simple competition between these two proteins for barbed end binding. Our analysis suggests that presence of twinfilin at the barbed end might promote association of cofilin to filament sides. Interestingly we discovered that pair-wise interactions can successfully predict depolymerization dynamics in simultaneous presence of all three proteins. Taken together, our results provide novel insights on simultaneous interactions of profilin, cofilin and twinfilin with filament barbed ends using a promising approach combining theory and experiments. We believe this approach will have far-reaching implications for elucidating underlying principles which govern multicomponent protein dynamics across a range of biological processes.

**Results**

**Quantitative characterization of effects of individual proteins on barbed end depolymerization**

Previous studies have characterized the effects of profilin, cofilin and twinfilin individually on barbed-end depolymerization of ADP-$P_i$ filaments (16, 21, 23, 28). To eliminate any inconsistencies due to use of different biochemical conditions and TIRF approaches amongst these studies, we first carried out a side-by-side quantitative characterization of each of these proteins using microfluidics-assisted TIRF microscopy (mf-TIRF) (Fig. 1a,b). Actin filaments were grown by exposing coverslip-anchored spectrin-actin seeds to a solution containing fluorescently-labelled actin monomers and profilin (22) (Fig. 1d). These filaments were all anchored at their pointed ends and their barbed ends remained free, thus allowing unambiguous recording of barbed-end depolymerization of hundreds of filaments. Filaments were maintained in ADP-$P_i$ state throughout the experiment by supplementing the TIRF buffer with 50 mM $P_i$ (23) (see methods). Using this setup, we first measured the depolymerization rate of filament barbed ends in presence of each of these depolymerases individually. Consistent with earlier studies, all three proteins increased barbed-end depolymerization of ADP-$P_i$ filaments in a concentration-dependent fashion (Fig. 1f-h).

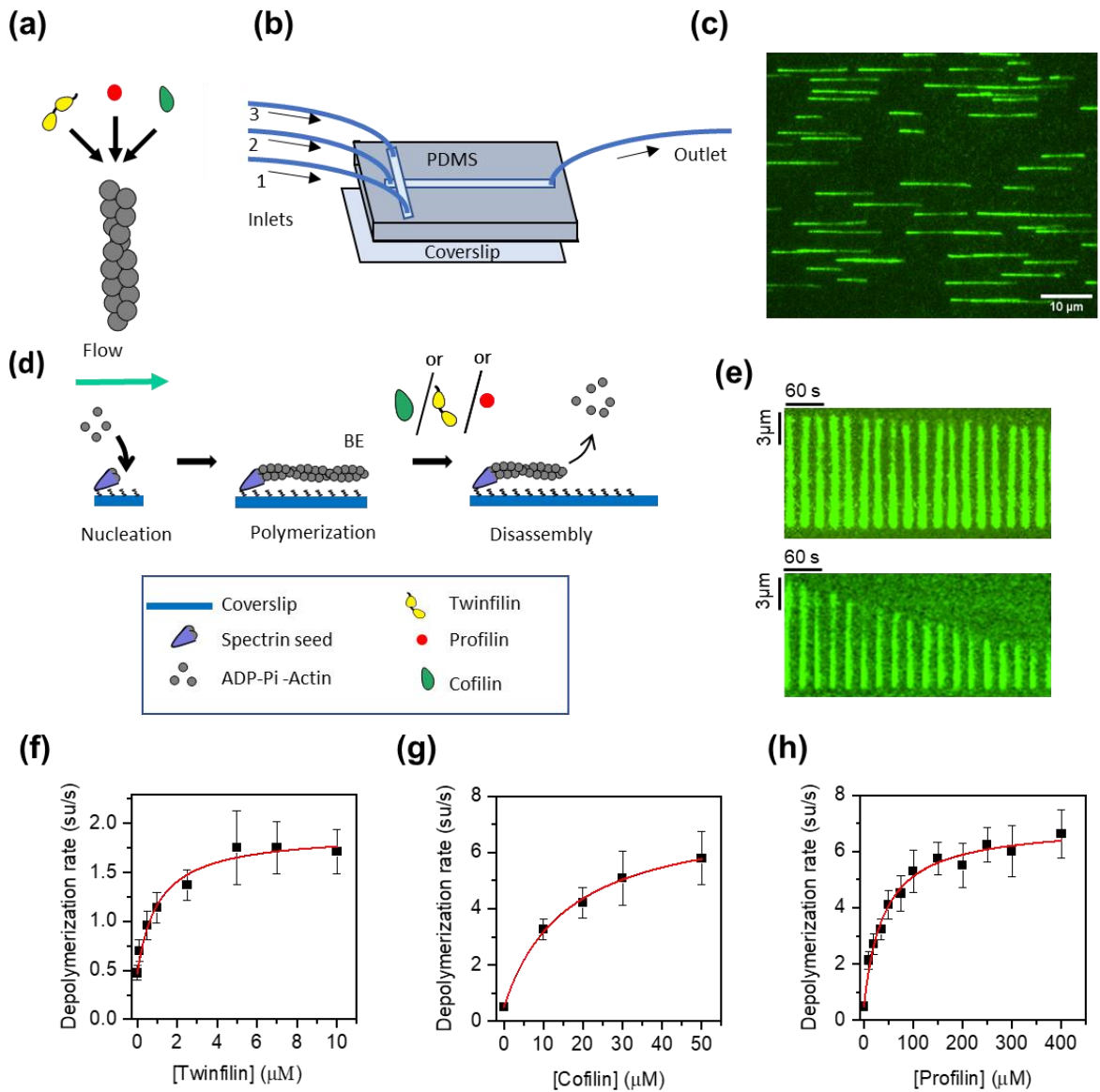

**Fig. 1: Cofilin, profilin and twinfilin individually accelerate depolymerization of ADP-P$_i$ actin filament barbed ends. (a)** Schematic representation of three depolymerases individually acting at the barbed end. **(b)** Schematic representation of microfluidics setup. **(c)** An example of field of view of actin filaments in mf-TIRF microscopy **(d)** Experimental strategy for measuring barbed-end depolymerization of ADP-P$_i$ filaments in presence of cofilin, profilin or twinfilin individually. Actin filaments with free barbed end were polymerized from coverslip-anchored spectrin-actin seeds by introducing 1 µM G-actin (15% Alexa-488 labeled) and 4 µM profilin in modified TIRF buffer. These filaments were then exposed to a flow containing cofilin, twinfilin, or profilin and the rate of depolymerization at their barbed ends (BE) was monitored. **(e)** Representative kymographs of Alexa-488-labeled actin filament (green) depolymerizing in absence (buffer, control) (top) and in presence of 100 µM profilin (bottom). **(f)** Rates (mean ± SD) of barbed-end depolymerization as a function of mTwf1 concentration. Number of filaments analyzed for each concentration (left to right): 62, 64, 62, 59, 67, 62, 64 and 67. A fit (line) of a simple kinetic model of depolymerization to the data is shown (see Equation (2) and SI). **(g)** Rate (mean ± SD) of barbed-end depolymerization as a function of cofilin concentration. Number of filaments analyzed for each concentration (left to right): 51,144,142,136, and 79. A fit to the model is shown (see SI). **(h)** Rates (mean ± SD) of barbed end depolymerization as a function of cofilin concentration. Number of filaments analyzed for each concentration (left to right): 66, 79, 79, 73, 72, 56, 82, 93, 98, 87, 72 and 63. A fit to the model is shown (see SI). Experiments in (d), (e), and (f) were performed three times and yielded similar results. The data shown are from one experiment each. Here su/s refers to subunits/s.

To understand this data, we developed a simple thermodynamic model of actin filament depolymerization (Supplementary Figure 1a). A key assumption of this model is that the timescales associated with binding and unbinding of various depolymerases to actin barbed ends are much faster than the rate of depolymerization. Similar thermodynamic models have been earlier applied to understanding mechanism regulating transcription (43, 44). In this model, a single protein molecule *P* binds to the barbed end of a bare actin filament. As a result, the barbed can either be in a bare (free) state or a protein-bound state. Using statistical mechanics, we can obtain the equilibrium weights of these two states respectively (43), which are dictated by the protein's dissociation constant ($K_{D,P}$) and its concentration ($C_P$) in the solution. Rates of depolymerization in each of these states are given by $d_0$ and $d_1$ respectively. The average depolymerization rate ($D_P$) of a filament can then be written as

$$D_P = d_0 p_0 + d_1 p_1, \quad (1)$$

where $p_0$ is the probability of the filament end being in the bare state and $p_1$ is the probability of the filament end being in protein-bound state (see SI for details). After a bit of algebra, the depolymerization rate can be rewritten as

$$D_P = \frac{d_0 K_{D,P} + d_1 C_P}{C_P + K_{D,P}}. \quad (2)$$

We use Equation 2 to compare our model to experimental data. Our model fits the experimental data well (Fig. 1f-h) and allows extraction of depolymerization rates in the bare state ($d_0$), protein bound state ($d_1$), and dissociation constant ($K_{D,P}$) for each of these proteins separately (see Table 1 below for fitted parameters). The rate of depolymerization from the bare barbed end is given by $d_0 = 0.50 \pm 0.08$ su/s, as extracted from experimental data. As expected from our model, the experimentally observed depolymerization rates increased with the bulk concentration and eventually saturated. At saturating concentration, the barbed end is almost always occupied by a protein molecule. The quantitative characterization carried out in this section paves the way for studying the combined effects of multiple proteins on barbed-end depolymerization.

**Table 1: Depolymerization rates and dissociation constants for twinfilin, profilin, and cofilin**

| Protein | $K_D$ (µM) | $d_1$ (su/s) |
|---|---|---|
| Twinfilin | 0.77 ± 0.32 | 1.93 ± 0.18 |
| Profilin | 18.7 ± 4.8 | 6.6 ± 0.4 |
| Cofilin | 19 ± 6 | 8.3 ± 1.0 |

Table 1: Depolymerization rates and dissociation constants for twinfilin, profilin, and cofilin obtained from fitting Equation (2) to experimental data in Fig. 1f-h where $K_D$ is the dissociation constant and $d_1$ is the depolymerization rate when twinfilin/profilin/cofilin is bound to barbed end.

**Uncovering multi-protein barbed-end depolymerization using competitive and simultaneous binding models**

We have thus far quantified the effects of individual proteins. In living cells however, multiple proteins simultaneously regulate actin dynamics. Proteins targeting the same site on a filament can either bind competitively (i.e., mutually exclusively) or simultaneously. We developed two broad classes of thermodynamic models to discriminate between these two possible modes of interaction i.e., competitive or simultaneous binding. In the competitive binding model, two proteins bind the barbed end in a mutually exclusive manner, whereby only one of them occupies the barbed end at a given moment. In contrast, in the simultaneous binding model, two proteins can simultaneously occupy the same barbed end. Below, we discuss the two models in greater detail.

When two proteins A and B bind the barbed end competitively (following the competitive binding model), the barbed end can exist in three different states – free, protein A bound, or protein B bound as shown in Fig. 2a. The statistical weights of these different states in equilibrium are governed by the dissociation constants of the two proteins and their concentrations respectively (43). Each of these three states is characterized by its distinct depolymerization rate – $d_0$ (free), $d_{1,A}$ (protein A bound) and $d_{1,B}$ (protein B bound). Switching between these states results in an average depolymerization rate given by

$$D_{AB} = \frac{d_0 K_{D,A} K_{D,B} + d_{1,A} C_A K_{D,B} + d_{1,B} C_B K_{D,A}}{K_{D,A} K_{D,B} + C_A K_{D,B} + C_B K_{D,A}}, \quad (3)$$

where $C_A$ and $C_B$ are concentrations of the two proteins, and $K_{D,A/B}$ is the dissociation constant of protein A or B at the barbed end (see SI for details). The competitive binding model exhibits a complex concentration-dependent behavior as shown in Fig. 2c. At low concentrations, the average depolymerization rate in simultaneous presence of both the proteins is higher than when either of the proteins is present alone. In contrast, at higher concentrations the average depolymerization rate with two proteins present lies somewhere in between the depolymerization rates of individual proteins.

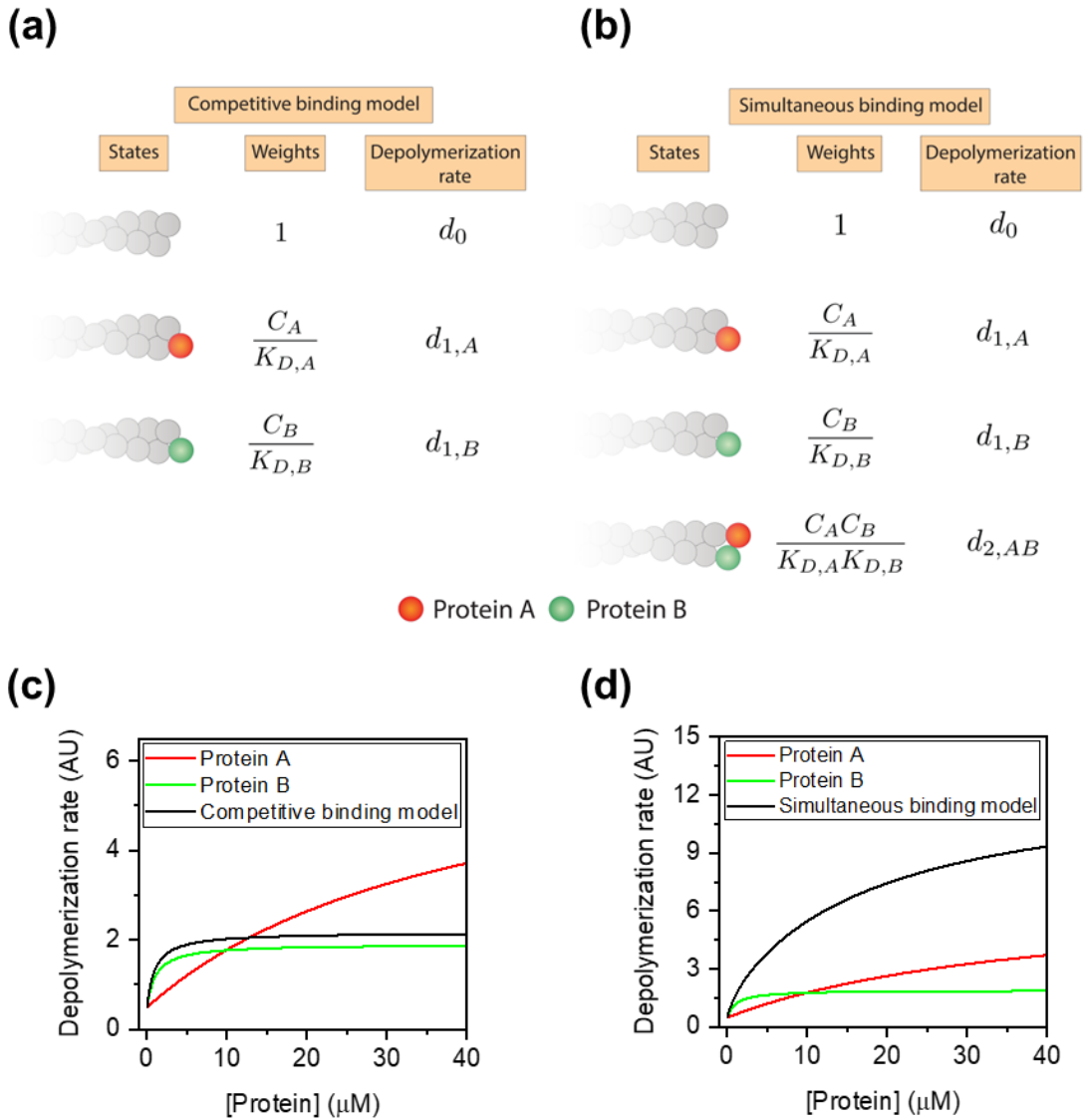

**Fig. 2: (a)** Thermodynamic model for competitive interaction of two proteins at the barbed end. At equilibrium, each of the microstates have a corresponding statistical weight (43). These weights are governed by concentration ($C_{A,B}$) and dissociation constant ($K_{D,A/B}$) of the proteins. The rate of depolymerization from the three states are $d_0$ (free), $d_{1,A}$ (protein A bound) and $d_{1,B}$ (protein B bound) respectively. **(b)** Thermodynamic model for simultaneous interaction of two proteins at the barbed end. Each of the microstates have a corresponding equilibrium statistical weight (43). These weights are governed by concentration ($C_{A,B}$) and dissociation constant ($K_{D,A/B}$) of the proteins. The rate of depolymerization from the four states are $d_0$ (free), $d_{1,A}$ (protein A bound), $d_{1,B}$ (protein B bound) and $d_{2,AB}$ (both protein A and B simultaneously bound). **(c)** Average rate of barbed-end depolymerization ($D_{AB}$) in presence of A alone (red), B

alone (green), and A and B together (black) as a function of their concentration for competitive binding model (see SI). **(d)** Average rate of barbed-end depolymerization ($D_{AB}$) in presence of A alone (red), B alone (green), and A and B together (black) as a function of their concentration for simultaneous binding model (see SI).

Next, we considered the second class of model, namely the simultaneous binding model (Fig. 2b). In contrast to the competitive binding model, the simultaneous binding model can lead to an additional barbed-end state when the two proteins A and B are both simultaneously bound to the barbed end. Consequently, the barbed end can now exist in four distinct states - free, protein A bound, protein B bound or both protein A and B bound. The rate of depolymerization from these different states are given by – $d_0$ (free), $d_{1,A}$ (protein A bound), $d_{1,B}$ (protein B bound) and $d_{2,AB}$ (both protein A and B simultaneously bound). The average depolymerization for this model is given by

$$D_{AB} = \frac{d_0 K_{D,A} K_{D,B} + d_{1,A} C_A K_{D,B} + d_{1,B} C_B K_{D,A} + d_{2,AB} C_A C_B}{K_{D,A} K_{D,B} + C_A K_{D,B} + C_B K_{D,A} + C_A C_B}, \qquad (4)$$

where $C_A$ and $C_B$ are concentrations of the two proteins, and $K_{D,A/B}$ is the dissociation constant of protein A or B at the barbed end. A key assumption of our model is that the dissociation constants of either of the proteins remain unaffected by the presence of the other protein. The simultaneous binding model predicts that the average depolymerization rate in the presence of both proteins will either be additive or super-additive as compared to the individual depolymerization rates of proteins as shown in Fig. 2d.

Overall, the competitive and cooperating models make distinct falsifiable predictions that can be tested experimentally and allow us to extract mechanistic insights into multicomponent depolymerization of actin, as we demonstrate in the ensuing sections.

**Profilin and twinfilin bind competitively to filament barbed ends**

We first asked how profilin and twinfilin together depolymerize actin filament barbed ends. X-ray diffraction structural studies indicate that twinfilin and profilin both bind to the barbed surface of actin monomers (40, 42, 45). Based on this, we hypothesized that these proteins bind in a mutually exclusive manner to filament barbed ends as well.

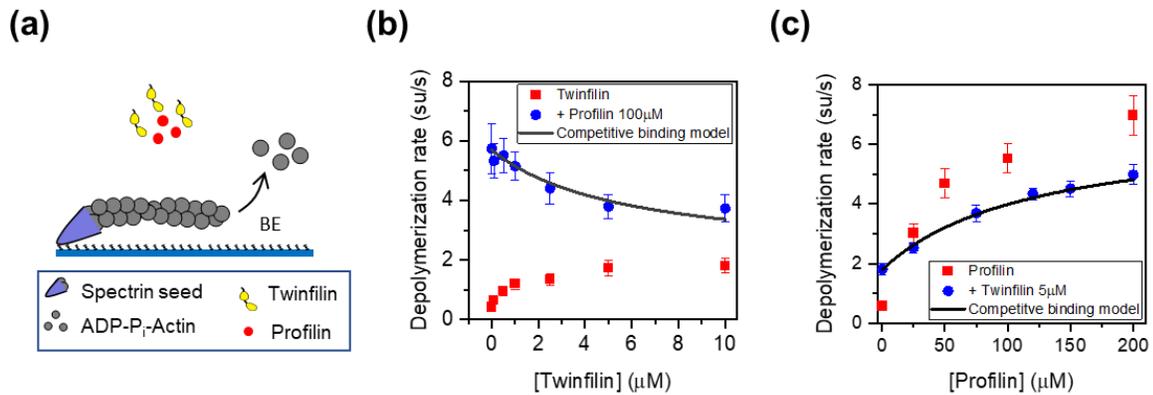

**Fig. 3: Profilin and twinfilin bind filament barbed ends competitively. (a)** Schematic depiction of the experimental strategy for measuring barbed-end depolymerization of ADP-$P_i$ actin filaments in simultaneous presence of twinfilin and profilin. Actin filaments with free barbed end were polymerized from coverslip-anchored spectrin-actin seeds by introducing 1 µM G-actin (15% Alexa-488 labeled) and 4 µM profilin in modified TIRF buffer containing 50 mM $P_i$. These filaments were then exposed to a flow containing twinfilin and profilin, and their rate of barbed-end (BE) depolymerization was monitored. **(b)** Rates (mean ± SD) of barbed-end depolymerization in presence of range of concentrations of twinfilin alone (red symbols) or additionally supplemented with 100 µM profilin (blue symbols). Number of filaments analyzed for each concentration of twinfilin (red curve, left to right): 67, 54, 55, 52, 26, 37 and 42. Number of filaments analyzed for 100 µM profilin and twinfilin (blue curve, left to right): 63, 65, 64, 65, 63, 58 and 61. The experimental data was compared to predictions from competitive binding model of depolymerization (black curve) (see SI). **(c)** Rate (mean ± SD) of barbed-end depolymerization in presence of range of concentrations of profilin alone (red symbols) or additionally supplemented with 5 µM twinfilin (blue symbols). The number of filaments analyzed for each concentration of Profilin (red symbols, left to right): 32, 38, 61, 50 and 61. Number of filaments analyzed for 5 µM twinfilin and profilin (blue symbols, left to right): 60, 60, 60, 60, 60 and 60. The experimental data was compared to predictions from competitive binding model of depolymerization (black curve) (see SI).

Similar to earlier experiments, free barbed ends of ADP-$P_i$ actin filaments were exposed to a solution containing both profilin (human profilin-1) and twinfilin (mouse twinfilin-1) (Fig. 3a). We first systematically tuned the concentration of twinfilin while keeping profilin's concentration fixed and compared the recorded depolymerization rates with that of twinfilin alone. In absence of profilin, increasing concentrations of twinfilin led to a monotonic increase followed by a saturation at high concentrations (~5 µM). In sharp contrast, addition of profilin qualitatively altered the depolymerization behavior; increasing concentrations of twinfilin (in presence of profilin) caused a monotonic decrease (Fig. 3b). These depolymerization rates were then compared to predictions from our competitive binding model using Equation. 3 (see supp. info). The model prediction matched the experimental data well, thereby validating our hypothesis that profilin and twinfilin bind to the filament barbed end in a mutually competitive manner (Fig. 3b). To further validate the competitive binding model, we systematically varied the concentration of profilin keeping twinfilin concentration fixed (Fig. 3c). We found that the experimental data were once again consistent with the predictions of the competitive binding model. Interestingly, as profilin concentration was tuned, presence of twinfilin did not alter the qualitative behavior of the observed rate of depolymerization. However quantitatively, the observed depolymerization rate in presence of both proteins always remained lower than of profilin alone.

**Profilin and cofilin can simultaneously occupy the same filament barbed end**

After validating profilin and twinfilin's competitive binding to barbed ends, we decided to investigate how cofilin and profilin together impact barbed end depolymerization. Since twinfilin and cofilin belong to the same protein family, we hypothesized that they would bind the filament barbed end similarly. Therefore, we expected that similar to twinfilin, cofilin would also compete with profilin in a mutually exclusive fashion for the filament barbed ends. Indeed, earlier studies have shown that they both bind the barbed surface on actin monomers thereby indicating that they might bind filament barbed ends competitively (40-42, 46). Nevertheless, a direct demonstration has remained elusive. To this end, we decided to experimentally characterize filament depolymerization rates in the presence of these two proteins and then challenge the experimental results against predictions of our competitive binding model.

We exposed free barbed ends of ADP-$P_i$ actin filaments to a solution containing either profilin and cofilin alone or together (Fig. 4a). Keeping the concentration of profilin fixed, we first tuned the concentration of cofilin and compared the recorded depolymerization rates with that of cofilin alone (Fig. 4b). In absence of profilin, increasing concentrations of cofilin led to a monotonic increase in depolymerization rate. However, the addition of profilin led to an overall increase in average depolymerization rates with increasing concentrations of cofilin.

We then compared these experimental findings with predictions from our competitive binding model (Equation. 3). Contrary to our expectation, the competitive binding model failed to capture the experimental data, thereby falsifying our hypothesis that profilin and cofilin bind filament ends mutually exclusively (Fig. 4b). The experimentally measured average depolymerization rates were consistently greater than the depolymerization rates predicted by the competitive model across the probed concentration range. A similar behavior was observed when cofilin concentration was kept constant and profilin concentration was varied (Fig 4c).

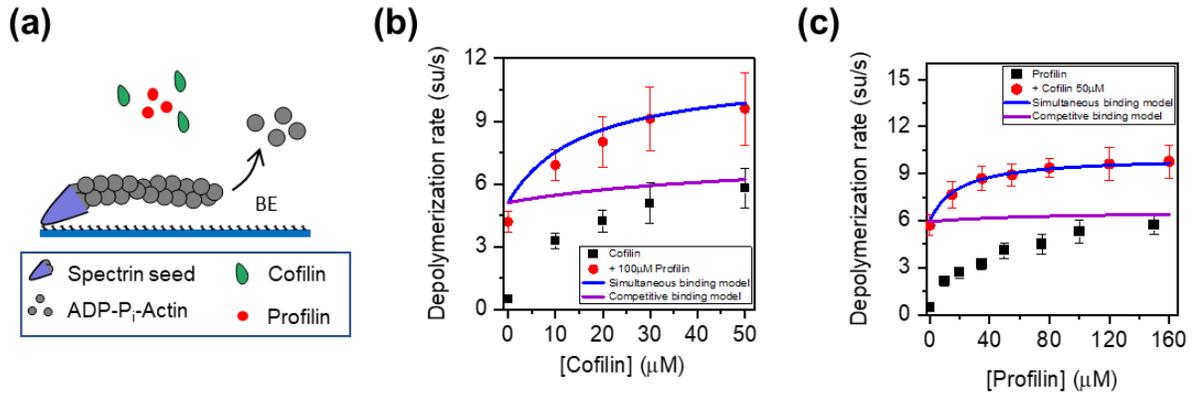

**Fig. 4: Competitive binding model fails to explain simultaneous interaction of profilin and cofilin with barbed ends. (a)** Schematic representation of the experimental strategy for measuring barbed end depolymerization of ADP-$P_i$ actin filaments in presence of cofilin and profilin. Actin filaments with free barbed end were polymerized from coverslip anchored spectrin-actin seeds by introducing 1 µM G-actin (15% Alexa-488 labeled) and 4 µM profilin in modified TIRF buffer containing 50 mM $P_i$. These filaments were then exposed to a flow containing cofilin and profilin, and the rate of depolymerization at their barbed ends (BE) was monitored. **(b)** Rates (mean ± SD) of barbed end depolymerization in presence of cofilin alone (black symbols) or in presence of 100 µM profilin (red symbols). Number of filaments analyzed for each concentration of cofilin (black symbols, left to right): 51, 144, 142, 136 and 79. Number of filaments analyzed for 100 µM profilin and cofilin (blue symbols, left to right): 106, 129, 151, 104 and 91. The experimental data was compared to predictions from competitive (purple) and simultaneous (blue) binding models (see SI). **(c)** Rates (mean ± SD) of barbed end depolymerization for ADP-$P_i$ filaments in presence of profilin alone (black symbols) or in presence of 50 µM cofilin (red symbols). The number of filaments analyzed for each concentration of profilin (black symbols, left to right): 66, 79, 79, 73, 72, 56, 82, and 93. Number of filaments analyzed for 50 µM cofilin and profilin (red symbols, left to right): 58, 63, 60, 60, 60, 61, and 92. The experimental data was compared to predictions from competitive (purple) and simultaneous (blue) binding models (see SI).

We suspected the possibility of simultaneous binding of profilin and cofilin at the barbed end, which could potentially lead to such high experimentally observed depolymerization rates. To test this possibility, we employed the simultaneous binding model (Equation. 4), as defined in the earlier section. Unlike the competitive binding model where the filament barbed end could exist in three distinct states, namely free, profilin-bound and cofilin-bound, the simultaneous binding model entails an additional fourth state where both cofilin and profilin are simultaneously bound to the barbed end. This fourth state is characterized by its own distinct depolymerization rate. *A priori*, this rate cannot be determined from experiments characterizing the depolymerization rates of individual proteins (Fig. 1d-f). Hence in order to test this model, we treated the depolymerization rate from the fourth state as a free parameter in our model. By tuning this depolymerization rate, we found that the simultaneous binding model indeed captured the experimentally measured depolymerization rates (Fig. 4b,c). Interestingly, this theory-experiment dialog demonstrates that profilin and cofilin can occupy the barbed end simultaneously. Moreover, when present together, the resulting depolymerization rate is higher than that of individual proteins. In summary, we showed that cofilin and profilin can simultaneously occupy the actin filament barbed end thereby enhancing the rate of depolymerization..

**Barbed-end bound twinfilin promotes association of cofilin**

After validating that while profilin competes with twinfilin and it can simultaneously occupy the same barbed end with cofilin, we then asked how cofilin and twinfilin together impact barbed-end depolymerization. Since twinfilin and cofilin both contain ADF/Cofilin homology domains, we hypothesized that these proteins would bind filament barbed ends in a mutually exclusive manner. To test this hypothesis, we measured barbed-end depolymerization of ADP-$P_i$ filaments in presence of these two proteins. Keeping the concentration of twinfilin fixed, we tuned the concentration of cofilin and compared the recorded depolymerization rates with that of cofilin alone (Fig. 5a,b). In absence of twinfilin, increasing concentrations of cofilin led to a monotonic increase in depolymerization rate. In sharp contrast, addition of twinfilin qualitatively altered the depolymerization behavior; increasing concentrations of cofilin (in presence of twinfilin) caused a monotonic decrease. Simultaneous presence of cofilin and twinfilin led to a significant reduction in the rate of depolymerization to values below 1 su/s (Fig. 5b). A similar behavior was observed when twinfilin concentration was kept constant and cofilin concentration was varied (Fig. 5c). A systematic comparison between experimental data with predictions from the competitive binding model falsified our hypothesis. Notably, while the model predicted an increase in average depolymerization rate as a function of cofilin concentration (at fixed twinfilin concentration), the experimental data showed the exact opposite trend. Next, we considered the possibility of twinfilin and cofilin simultaneously occupying the barbed end. This model also failed to explain the data. Taken together, we find that the experimental observations are inconsistent with both of our simple single-site barbed end binding models (Supplementary Figure 2a, 2b). This implies that barbed-end interactions of these two proteins (when simultaneously present) might be more complex.

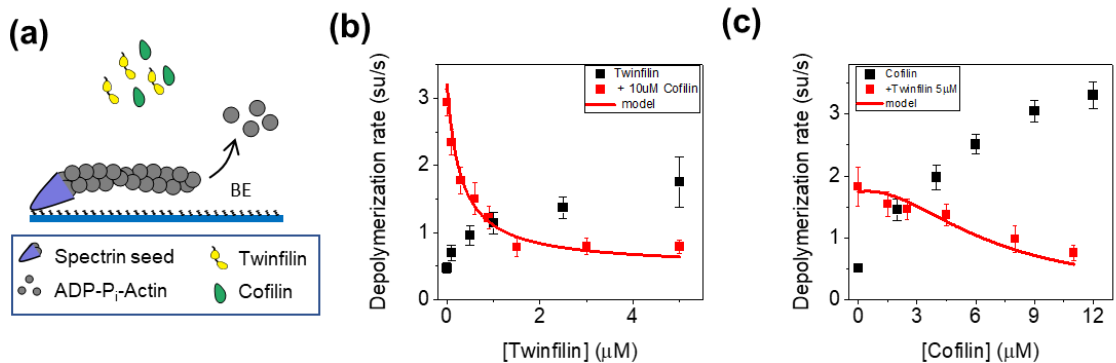

**Fig. 5: Twinfilin and cofilin bind filament barbed ends competitively. (a)** Schematic depiction of the experimental strategy for measuring barbed-end depolymerization of ADP-$P_i$ actin filaments in presence of twinfilin and cofilin. Actin filaments with free barbed ends were polymerized from coverslip-anchored spectrin-actin seeds by introducing 1 µM G-actin (15% Alexa-488 labeled) and 4 µM profilin in modified TIRF buffer containing 50 mM $P_i$. These filaments were then exposed to a flow containing twinfilin and cofilin, and the rate of depolymerization at their barbed ends (BE) was monitored. **(b)** Rates (mean ± SD) of barbed end depolymerization in presence of twinfilin alone (black symbols) or additionally supplemented with 10 µM cofilin (red symbols). Number of filaments analyzed for each concentration of twinfilin (black symbols, left to right): 62, 64, 62, 59, 67, and 62. Number of filaments analyzed for twinfilin and 10 µM cofilin (red symbols, left to right): 61, 62, 52, 63, 60, 45, 50 and 59. The experimental data was compared to predictions from the revised competitive (red) model (Equation 3, see SI). **(c)** Rates (mean ± SD) of barbed end depolymerization in presence of cofilin alone (black symbols) or additionally supplemented with 5 µM twinfilin (red symbols). Number of filaments analyzed for each concentration of cofilin (black symbols, left to right): 40, 60, 60, 60, 60, 60, and 60. Number of filaments analyzed for 5 µM twinfilin and cofilin (red

A recent study showed that binding of cofilin to the sides of ADP-actin filaments leads to a dramatic reduction in barbed-end depolymerization to rates lower than in control reactions (28). Although, side binding of cofilin has so far mainly been studied in the context of ADP-actin filaments, acceleration of phosphate release by cofilin suggests that it can also interact (albeit weakly) with sides of ADP-$P_i$ actin filaments (47, 48). We therefore asked if considering side-binding of cofilin in addition to barbed end binding can explain our experimental findings. Since we did not observe a reduction in barbed end depolymerization even at high cofilin concentrations (Fig. 1), we wondered if the added presence of twinfilin might stabilize the weak-interactions of cofilin with sides of ADP-$P_i$ actin filaments, thereby reducing the depolymerization rate.

Taking into account these possibilities, we expanded our barbed-end competitive model to include side binding of cofilin, aided by barbed-end presence of twinfilin. Specifically, we considered side-binding of one cofilin molecule per actin protofilament (near the barbed end). In this revised competitive model, we used the cofilin side binding dissociation constant $K_{D,SC}$= 181µM for newly assembled actin filaments (i.e., ADP-$P_i$-F-actin) as determined in a previous single-molecule study (49) and kept ω, the cooperativity between side binding of cofilin and barbed end-bound twinfilin as a free parameter (Supplementary Figure 1b). To test this model, we fit it to our experimental data where we varied twinfilin at fixed cofilin concentration (Fig. 5b). We found that this model captured experimental data well and allowed us to infer the free parameter ω. Using the inferred parameter, we predicted the effect of varying cofilin concentration at fixed twinfilin concentration (Fig. 5c). The average depolymerization rate as predicted by the model matches the data well, thereby validating our revised competitive model. Taken together, our unique approach has enabled us to uncover a potentially new mechanism in which presence of twinfilin at the filament barbed end promotes the association of cofilin to filament sides which results in a dramatic reduction of barbed end depolymerization.

**Pair-wise interactions are sufficient to explain three-protein interactions**

Following pair-wise investigation of profilin, cofilin, and twinfilin, we asked if we could leverage our learnings from their pair-wise interactions to predict depolymerization dynamics of the barbed end when all three proteins were present simultaneously (Fig. 6). To this end, we constructed a model for the three-protein case by combining their pair-wise interactions (see Supplementary Figure 1c). This model makes specific predictions for the average depolymerization rate in the presence of all three proteins. In particular, we made two sets of predictions. First, we varied twinfilin concentration in presence of fixed concentration of profilin and cofilin. To our surprise, our model predicted a steeper drop in depolymerization rates upon addition of cofilin in comparison when only twinfilin and profilin were present (Fig. 6b). Second, we varied cofilin concentration in presence of fixed concentration of twinfilin and profilin. Once again, our model predicted a much steeper drop in average depolymerization upon addition of profilin in comparison to when only cofilin and twinfilin were present (Fig. 6c). We put these predictions to test using experiments. We find excellent agreement between the predictions and our experimental measurements. Taken together, our results show that pair-wise interactions alone are sufficient to predict multicomponent depolymerization dynamics resulting from three proteins.

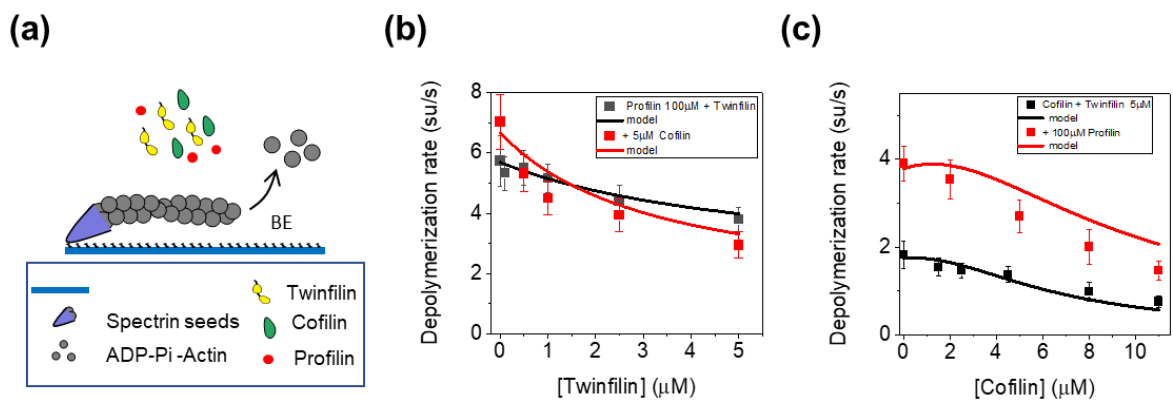

**Fig. 6: Three-protein model based on pairwise interactions fails to explain depolymerization dynamics in presence of profilin, twinfilin and cofilin. (a)** Schematic representation of the experimental strategy for measuring barbed-end depolymerization of ADP-$P_i$ actin filaments in presence of profilin, twinfilin, and cofilin. Actin filaments with free barbed ends were polymerized from coverslip anchored spectrin-actin seeds by introducing 1 µM G-actin (15% Alexa-488 labeled) and 4 µM profilin in modified TIRF buffer containing 50 mM $P_i$. These filaments were then exposed to a flow containing profilin, twinfilin, and cofilin and the rate of depolymerization at their barbed ends (BE) was monitored. **(b)** Rates (mean ± SD) of barbed-end depolymerization in presence of 100 µM profilin and twinfilin (black symbols) or additionally supplemented with 5 µM cofilin (red symbols). Number of filaments analyzed

for each concentration of 100 µM profilin and twinfilin (black symbols, left to right): 63, 65, 64, 65, 63, and 58. Number of filaments analyzed for 100 µM profilin, 5 µM cofilin, and twinfilin (red symbols, left to right): 76, 60, 60, 61, and 62. The experimental data was compared to predictions from two-protein competitive (black) model for 100 µM profilin and twinfilin and three protein model (red) for 100 µM profilin, 5 µM cofilin, and twinfilin (see SI). **(c)** Rates (mean ± SD) of barbed-end depolymerization in presence of 5 µM twinfilin and cofilin (black symbols) or additionally supplemented with 100 µM profilin (red symbols). Number of filaments analyzed for each concentration of 5 µM twinfilin and cofilin (black symbols, left to right): 50, 51, 37, 63, 42 and 38. Number of filaments analyzed for 5 µM twinfilin, 100 µM profilin, and cofilin (red symbols, left to right): 60, 63, 61, 58, and 42. The experimental data was compared to predictions from two-protein competitive (black) model for 5 µM twinfilin and cofilin and three protein model (red) for 5 µM twinfilin, 100 µM profilin, and cofilin (see SI).

**Discussion**

A large array of proteins interact with actin filaments to facilitate intracellular actin assembly and remodeling (1). Majority of these interactions occur at the barbed end of the actin filament (4). As a result, these proteins can either bind filament barbed ends in a mutually exclusive manner or simultaneously associate with filament barbed ends. However, directly visualizing multiple proteins simultaneously interacting with the barbed end is technically challenging. Here we have presented an alternate approach that combines experimental measurements with predictive theoretical modelling to fill this gap. We show that high throughput measurements coupled with theory can help uncover features of multicomponent dynamics that would otherwise remain intractable. We believe our approach is especially useful when studying multicomponent dynamics of proteins that bind filament very transiently (<<1 s), as directly visualizing these short-lived interactions using classical methods like multicolor single-molecule imaging is not possible.

In particular, we looked at three proteins – profilin, cofilin and twinfilin all of which individually depolymerize barbed ends of newly assembled (ADP-$P_i$) actin filaments. To dissect how they simultaneously regulate actin depolymerization, we experimentally measured their effects individually, in pairs and all together, and then challenged these measurements against predictions of our theoretical models.

Cofilin and profilin compete for binding actin monomers (41). We therefore hypothesized that the two proteins would also bind filament barbed ends in a mutually exclusive manner. To our surprise, we found that cofilin and profilin can simultaneously co-occupy filament barbed ends. Our analysis showed that the competitive binding model failed to explain the experimentally observed depolymerization rates when both cofilin and profilin are present. We instead found that the simultaneous binding model captured the experimentally observed depolymerization rates and therefore, supports the alternative possibility that cofilin and profilin can simultaneously occupy the actin filament barbed end. How do we structurally rationalize the simultaneous binding of these two proteins to filament barbed ends? To do this, we used co-crystal structures of human profilin bound to an actin monomer (42) and ADF/cofilin homology domain bound to an actin monomer (40). Both of these were docked on the two terminal actin subunits of an ADP-$P_i$ actin filament (50). In agreement with Courtemanche and Pollard (16), we found that while profilin can bind the terminal actin subunit (n) with minimal clashes, it clashes directly with the terminal actin subunit (n) when docked on the penultimate subunit (n-1). This suggests that profilin's interactions with the filament barbed end might originate solely from its interaction with the terminal barbed-end subunit. Upon docking the Cofilin/ADF homology domain on the actin filament, we however found that it could comfortably interact with both the terminal and penultimate actin subunits (n and n-1) at the barbed end. Taken together, our structural analysis suggests that the most likely explanation for our findings is that profilin binds to the terminal actin subunit and cofilin preferentially to the penultimate one, thus providing a pathway for them to be able to occupy the filament barbed end simultaneously. As a result, simultaneous presence of profilin and cofilin destabilizes both the subunits at the barbed end causing fastest ever depolymerization rates measured for ADP-$P_i$ barbed ends (Fig. 4b-c).

Since twinfilin and cofilin are both members of the ADF/Cofilin family of proteins, twinfilin would also be expected to simultaneously bind barbed ends with profilin. Surprisingly, our analysis showed that unlike cofilin and profilin, twinfilin and profilin bind barbed ends in a mutually exclusive manner. While twinfilin contains two connected ADF-Homology domains connected via a short peptide linker, cofilin only contains a single ADF-Homology domain. In absence of X-ray/cryo-EM structure of twinfilin-bound filament barbed ends, the mechanism of twinfilin-mediated depolymerization is less settled. Two separate modes of binding have been proposed. Twinfilin can either bind both the terminal and the penultimate actin subunits via its two ADF/Cofilin homology domains (45) or bind to only one of the two subunits at the barbed ends (26). In the former configuration, twinfilin's presence would sterically hinder profilin's association to both the terminal actin subunits. In the latter configuration, one of the ADF/cofilin homology domain of twinfilin would interact with either the terminal or the penultimate subunit and the second ADF/Cofilin domain would interact with the side of the actin filament (26). In this scheme, the barbed end subunit not bound to twinfilin will still be able to interact with profilin. Based on these two alternative proposals, we developed kinetic models that made specific predictions about the observed depolymerization rate as a function of profilin and twinfilin concentration. Our experimental results favor the first proposal over the second i.e., twinfilin associates with both the barbed-end subunits. Taken together, twinfilin's presence at the filament barbed ends prevents profilin's interactions with both actin subunits at the filament barbed end. Importantly, we note that these results also demonstrate the power of our theory-experiment approach – using profilin as a probe, our analysis has provided novel insights into twinfilin's binding to and depolymerization of barbed ends.

Since twinfilin and cofilin both contain ADF/Cofilin homology domains, we hypothesized that these proteins would bind filament barbed ends in a mutually exclusive manner. However, to our surprise, we found that the depolymerization dynamics for these two proteins could not be explained by our simple single-site competitive binding model. Specifically, we found that experimentally measured depolymerization rates in simultaneous presence of cofilin and twinfilin were always much slower than predicted by the model. What might explain this? Cofilin readily decorates the sides of ADP filaments. At saturating concentrations, complete side decoration of actin filaments by cofilin leads to a 10-fold reduction in barbed-end depolymerization compared to bare actin filaments (28, 47). However, we did not see a reduction in depolymerization rate of ADP-$P_i$ filaments even at high cofilin concentrations suggesting that cofilin alone might not stably decorate sides of actin filaments. We therefore wondered if the presence of twinfilin at the barbed end could stabilize interactions of weak interactions of cofilin with sides of ADP-Pi actin filaments. Indeed, our revised competitive model which includes cofilin side binding and its cooperative interactions with barbed-end bound twinfilin. Experimentally, it is extremely challenging to determine whether cofilin-mediated barbed end stabilization requires cofilin's side binding only to the terminal actin subunits or to a much larger stretch of the actin filament. Our analysis has revealed that in presence of twinfilin, cofilin's association to only the last two barbed end subunits might be sufficient to dramatically reduce barbed end depolymerization.

After deciphering principles that govern two-protein interactions, we asked if these pair-wise interactions are sufficient to describe the dynamics of the system when all three proteins are simultaneously present. To this end, we constructed a model for three-protein case explicitly incorporating the three sets of pair-wise interactions. This model made specific predictions for how the average depolymerization rates would change as we tuned the concentration of one of the proteins while we kept the other two fixed. In this case, pair-wise interactions were sufficient to explain our results from three-protein experiments indicating an absence of higher order interactions in this system unlike in other complex systems higher order interactions play a significant role in dictating multicomponent interactions (51, 52).

Do the depolymerization mechanisms uncovered have *in vivo* relevance? Two questions come up. Firstly, filaments at the leading edge exhibit lifetimes of only a few seconds (53), which is about two orders of magnitude faster than timescales of $P_i$ release (0.002 $s^{-1}$) (48, 54). This discrepancy suggests that it is possible that filaments *in vivo* might entirely bypass the phosphate release step and get depolymerized in their ADP-$P_i$ state possibly by multicomponent mechanisms described here. Secondly, while the experiments in this study were conducted in the absence of actin monomers, cytoplasm is thought to contain up to 100 µM monomeric actin, mostly bound to either profilin or thymosin β4. As a result, availability of free profilin in the cytosol remains unclear. A study from the Pollard lab found that up to 40% of total profilin might be free of G-actin in the acanthamoeba cytosol (55). Further, a recent study from Bieling lab suggested that about 25 to 50 µM free profilin might be present in the cytoplasm of vertebrate lymphocytes and dendritic cells (56). These estimates of free profilin assume that about 50% of total cytosolic actin remains in the filamentous state and the other half in monomeric state (30, 57). Further, it is important to recognize that to gain mechanistic insights into how these regulators act at the barbed end, it is preferable to work under simple conditions in which no interference occurs from the interaction of these proteins with actin monomers. It is thus an important step towards understanding more complex intracellular conditions.

In summary, we implemented a bottom-up approach that builds on single protein measurements to shed light on emergent multi-protein behavior. Single protein measurements allow for the extraction of various rate constants, which can be utilized to build predictive models of multi-protein regulation of actin dynamics. These models can be falsified through a rigorous comparison of their predictions with experimental data. While we used this approach to shed new light on actin depolymerization by profilin, cofilin, and twinfilin, we believe this approach should be generally applicable to decipher of how living cells integrate activities of multiple proteins to regulate complex intracellular actin dynamics.

**Acknowledgements:** We thank Heidi Ulrichs for purifying profilin. S.C. is supported by the Ramalingaswami Re-entry Fellowship (BT/HRD/35/02/2006) from Department of Biotechnology, Government of India. S.S. is supported by NIH NIGMS grant R35GM143050 and startup funds from Emory University.

**Methods**

**Purification and labeling of actin**

Rabbit skeletal muscle actin was purified from acetone powder generated from frozen ground hind leg muscle tissue of young rabbits (PelFreez, USA). Lyophilized acetone powder stored at −80°C was mechanically sheared in a coffee grinder, resuspended in G-buffer (5 mM Tris-HCl pH 7.5, 0.5 mM Dithiothreitol (DTT), 0.2 mM ATP and 0.1 mM $CaCl_2$), and cleared by centrifugation for 20 min at 50,000 × $g$. Supernatant was collected and further filtered with Whatman paper. Actin was then polymerized overnight at 4°C, slowly stirring, by the addition of 2 mM $MgCl_2$ and 50 mM NaCl to the filtrate. The next morning, NaCl powder was added to a final concentration of 0.6 M and stirring was continued for another 30 min at 4°C. Then, F-actin was pelleted by centrifugation for 150 min at 280,000 × $g$, the pellet was solubilized by dounce homogenization and dialyzed against G-buffer for 48 h at 4°C. Monomeric actin was then precleared at 435,000 × $g$ and loaded onto a Sephacryl S-200 16/60 gel-filtration column (Cytiva, USA) equilibrated in G-Buffer. Fractions containing actin were stored at 4°C.

To fluorescently label actin, G-actin was polymerized by dialyzing overnight against modified F-buffer (20 mM PIPES pH 6.9, 0.2 mM $CaCl_2$, 0.2 mM ATP, 100 mM KCl)(22). F-actin was incubated for 2 h at room temperature with a 5-fold molar excess of Alexa-488 NHS ester dye (Thermo Fisher Scientific, USA). F-actin was then pelleted by centrifugation at 450,000 × $g$ for 40 min at room temperature, and the pellet was resuspended in G-buffer, homogenized with a dounce and incubated on ice for 2 h to depolymerize the filaments. The monomeric actin was then re-polymerized on ice for 1 h by addition of 100 mM KCl and 1 mM $MgCl_2$. F-actin was once again pelleted by centrifugation for 40 min at 450,000 × $g$ at 4°C. The pellet was homogenized with a dounce and dialyzed overnight at 4°C against 1 L of G-buffer. The solution was precleared by centrifugation at 450,000 × $g$ for 40 min at 4°C. The supernatant was collected, and the concentration and labeling efficiency of actin was determined.

**Purification of twinfilin**

Mouse mTwf1 was expressed in *E. coli* BL21 (pRare). Cells were grown in Terrific Broth to log phase at 37°C. Expression was induced overnight at 18°C by addition of 1 mM IPTG. Cells were harvested by centrifugation at 11,200 × $g$ for 15 min and the cell pellets were stored at -80°C. For purification, frozen pellets were thawed and resuspended in 35 mL lysis buffer (50 mM sodium phosphate buffer pH 8, 20 mM imidazole, 300 mM NaCl, 1 mM DTT, 1 mM PMSF and protease inhibitors (pepstatin A, antipain, leupeptin, aprotinin, and chymostatin, 0.5 µM each)). Cells were lysed using a tip sonicator while kept on ice. The cell lysate was then centrifuged at 120,000 × $g$ for 45 min at 4°C. The supernatant was then incubated with 1 mL of Ni-NTA beads (Qiagen, USA) while rotating for 2 h at 4°C. The beads were then washed three times with the wash buffer (50 mM sodium phosphate buffer pH 8, 300 mM NaCl, 20 mM imidazole and 1 mM DTT). The beads were then transferred to a disposable column (Bio-Rad, USA). Protein was eluted using the elution buffer (50 mM phosphate buffer pH 8, 300 mM NaCl, 250 mM imidazole and 1 mM DTT). Fractions containing the protein were concentrated and loaded onto a size exclusion Superdex 75 Increase 10/300 column (Cytiva, USA) pre-equilibrated with 20 mM HEPES pH 7.5, 1 mM EDTA, 50 mM KCl and 1 mM DTT. Peak fractions were collected, concentrated, aliquoted, and flash-frozen in liquid $N_2$ and stored at -80°C.

**Purification of profilin**

Human profilin-1 was expressed in *E. coli* strain BL21 (pRare) to log phase in LB broth at 37°C and induced with 1 mM IPTG for 3 h at 37°C. Cells were then harvested by centrifugation at 15,000 × $g$ at 4°C and stored at -80°C. For purification, pellets were thawed and resuspended in 30 mL lysis buffer (50 mM Tris-HCl pH 8, 1 mM DTT, 1 mM PMSF protease inhibitors (0.5 µM each of pepstatin A, antipain, leupeptin, aprotinin, and chymostatin)) was added, and the solution was sonicated on ice by a tip sonicator. The lysate was centrifuged for 45 min at 120,000 × $g$ at 4°C. The supernatant was then passed over 20 ml of Poly-L-proline conjugated beads in a disposable column (Bio-Rad, USA). The beads were first washed at room temperature in wash buffer (10 mM Tris pH 8, 150 mM NaCl, 1 mM EDTA and 1 mM DTT) and then washed again with 2 column volumes of 10 mM Tris pH 8, 150 mM NaCl, 1 mM EDTA, 1 mM DTT and 3 M urea. Protein was then eluted with 5 column volumes of 10 mM Tris pH 8, 150 mM NaCl, 1 mM EDTA, 1 mM DTT and 8 M urea. Pooled and concentrated fractions were then dialyzed in 4 L of 2 mM Tris pH 8, 0.2 mM EGTA, 1 mM DTT, and 0.01% $NaN_3$ (dialysis buffer) for 4 h at 4°C. The dialysis buffer was replaced with fresh 4 L buffer and the dialysis was continued overnight at 4°C. The protein was centrifuged for 45 min at 450,000 × $g$ at 4°C, concentrated, aliquoted, flash frozen in liquid $N_2$ and stored at -80°C.

**Purification of Cofilin**

Human Cofilin-1 was expressed in *E.coli* BL21 DE3 cells. Cells were grown in Terrific Broth to log phase at 37ºC, and then expression was induced overnight at 18ºC by addition of 1 mM

IPTG. Cells were collected by centrifugation and pellets were stored at -80ºC. Frozen pellets were thawed and resuspended in lysis buffer (20 mM Tris pH 8.0, 50 mM NaCl, 1 mM DTT, and protease inhibitors (0.5 µM each of pepstatin A, antipain, leupeptin, aprotinin, and chymostatin). Cells were lysed with a tip sonicator while being kept on ice. The cell lysate was centrifuged at 150,000 × $g$ for 30 min at 4ºC. The supernatant was loaded on a 1 ml HisTrap HP Q column (GE Healthcare, Pittsburgh, PA), and the flow-through was collected and dialyzed against 20 mM HEPES pH 6.8, 25 mM NaCl, and 1 mM DTT. The dialyzed solution was then loaded on a 1 ml HisTrap SP FF column (GE Healthcare, Pittsburgh, PA) and eluted using a linear gradient of NaCl (20-500 mM). Fractions containing protein were concentrated, dialyzed against 20 mM Tris pH 8.0, 50 mM KCl, and 1 mM DTT, flash frozen in liquid $N_2$ and stored at -80ºC.

**Microfluidics-assisted TIRF (mf-TIRF) microscopy**

Actin filaments were first assembled in microfluidics-assisted TIRF (mf-TIRF) flow cells (21, 22). For all experiments, coverslips were first cleaned by sonication in Micro90 detergent for 20 min, followed by successive 20 min sonications in 1 M KOH, 1 M HCl and 200 proof ethanol for 20 min each. Washed coverslips were then stored in fresh 200 proof ethanol. Coverslips were then washed extensively with $H_2O$ and dried in an $N_2$ stream. These dried coverslips were coated with 2 mg/mL methoxy-poly (ethylene glycol) (mPEG)-silane MW 2,000 and 2 µg/mL biotin-PEG-silane MW 3,400 (Laysan Bio, USA) in 80% ethanol (pH 2.0) and incubated overnight at 70°C. A 40 µm high PDMS mold with 3 inlets and 1 outlet was mechanically clamped onto a PEG-Silane coated coverslip. The chamber was then connected to a Maesflo microfluidic flow-control system (Fluigent, France), rinsed with modified TIRF buffer (regular TIRF buffer supplemented with inorganic phosphate : 10 mM imidazole pH 7.4, 34.8 mM $K_2HPO_4$ and 15.2 mM $KH_2PO_4$, 1 mM $MgCl_2$, 1 mM EGTA, 0.2 mM ATP, 10 mM DTT, 1 mM DABCO) and incubated with 1% BSA and 10 µg/mL streptavidin in 20 mM HEPES pH 7.5, and 50 mM KCl for 5 min. Biotin spectrin-actin seeds were attached on the glass coverslip. Actin filaments with free barbed ends were then elongated by exposing the spectrin-actin seeds to a flow containing 1 µM G-actin (15% Alexa-488 labeled) and 4 µM profilin. These filaments were then exposed to profilin, cofilin and twinfilin (alone or together) in modified-TIRF buffer. Barbed end depolymerization of these filaments was monitored. All experiments were conducted at room temperature.

**Image acquisition and analysis**

Single-wavelength time-lapse TIRF imaging was performed on a Nikon-Ti2000 inverted microscope equipped with a 40 mW 488 nm Argon laser, a 60X TIRF-objective with a numerical aperture of 1.49 (Nikon Instruments Inc., USA) and an IXON LIFE 888 EMCCD camera (Andor Ixon, UK). One pixel was equivalent to 144 × 144 nm. Focus was maintained by the Perfect Focus system (Nikon Instruments Inc., Japan). Time-lapsed images were acquired every 10 s using Nikon Elements imaging software (Nikon Instruments Inc., Japan).

Images were analyzed in Fiji (58). Background subtraction was conducted using the rolling ball background subtraction algorithm (ball radius 5 pixels). For each condition, between 50 and 100 filaments were acquired across multiple fields of view. To determine the rate of depolymerization, the in-built kymograph plugin was used to draw kymographs of individual filaments. The kymograph slope was used to calculate barbed end depolymerization rate of each individual filament (assuming one actin subunit contributes 2.7 nm to filament length). Data analysis and curve fitting were carried out in Microcal Origin. All experiments were repeated multiple times and yielded similar results. Data shown are from one trial. The predictions for the models were calculated using custom-written code in MATLAB.

# Supplementary information

**Supplementary figure 1: Thermodynamic model of actin depolymerization in presence of regulatory proteins.**

**(a)** 

| States | Weights | Depolymerization rate |
|---|---|---|
| (free) | 1 | $d_0$ |
| (P bound) | $\dfrac{C_P}{K_{D,P}}$ | $d_1$ |

**(b)**

| States | Weights | Depolymerization rate |
|---|---|---|
| | 1 | $d_0$ |
| | $\dfrac{C_C}{K_{D,C}}$ | $d_{1,C}$ |
| | $\dfrac{C_T}{K_{D,T}}$ | $d_{1,T}$ |
| | $\dfrac{2C_C}{K_{D,SC}}$ | $d_0$ |
| | $\dfrac{2C_C^2}{K_{D,C}K_{D,SC}}$ | $d_{1,C}$ |
| | $\dfrac{2C_C C_T}{K_{D,T}K_{D,SC}}$ | $d_{1,T}$ |
| | $\dfrac{C_C^2}{K_{D,SC}^2}$ | $d_0$ |
| | $\dfrac{C_C^3}{K_{D,C}K_{D,SC}^2}$ | $d_{1,C}$ |
| | $\dfrac{\omega C_C^2 C_T}{K_{D,T}K_{D,SC}^2}$ | $d_{2,TC}$ |

● Cofilin  ● Twinfilin

**(c)**

| States | Weights | Depolymerization rate |
|---|---|---|
| | 1 | $d_0$ |
| | $\dfrac{C_C}{K_{D,C}}$ | $d_{1,C}$ |
| | $\dfrac{C_T}{K_{D,T}}$ | $d_{1,T}$ |
| | $\dfrac{C_P}{K_{D,P}}$ | $d_{1,P}$ |
| | $\dfrac{C_C C_P}{K_{D,C}K_{D,P}}$ | $d_{2,PC}$ |
| | $\dfrac{2C_C}{K_{D,SC}}$ | $d_0$ |
| | $\dfrac{2C_C^2}{K_{D,C}K_{D,SC}}$ | $d_{1,C}$ |
| | $\dfrac{2C_C C_T}{K_{D,T}K_{D,SC}}$ | $d_{1,T}$ |
| | $\dfrac{C_C^2}{K_{D,SC}^2}$ | $d_0$ |
| | $\dfrac{C_C^3}{K_{D,C}K_{D,SC}^2}$ | $d_{1,C}$ |
| | $\dfrac{\omega C_C^2 C_T}{K_{D,T}K_{D,SC}^2}$ | $d_{2,TC}$ |

● Profilin  ● Cofilin  ● Twinfilin

**(a)** Thermodynamic model for a single protein at the barbed end. Each of the thermodynamic states have a corresponding equilibrium statistical weight (43). These weights are governed by concentration ($C_{A,P}$) and dissociation constant ($K_{D,P}$) of the protein. The rate of depolymerization from the two states are $d_0$ (free), $d_1$ (protein P bound). **(b)** Thermodynamic model for interaction of cofilin and twinfilin at barbed ends. At equilibrium, each of the microstates have a corresponding statistical weight (43). These weights are governed by concentrations ($C_C$ and $C_T$) and dissociation constants ($K_{D,C}$, $K_{D,T}$, and $K_{D,SC}$) of cofilin and twinfilin. **(c)** Thermodynamic model for three proteins based on pair-wise interactions between profilin, cofilin, and twinfilin. At equilibrium, each of the microstates have a corresponding statistical weight (43). These weights are governed by concentrations ($C_P$, $C_C$ and $C_T$) and dissociation constants ($K_{D,P}$, $K_{D,C}$ $K_{D,SC}$, and $K_{D,T}$) of these three proteins respectively. The corresponding depolymerization rates are $d_0$ (free end), $d_{1,P}$ (profilin-bound), $d_{1,C}$ (cofilin-bound), $d_{1,T}$ (twinfilin-bound) $d_{2,PC}$ (simultaneously bound to profilin and cofilin), and $d_{2,TC}$ (simultaneous binding of cofilin to sides and of twinfilin to the barbed end).

**Supplementary figure 2: Falsifying competitive and simultaneously binding model for cofilin and twinfilin interaction at the barbed end.**

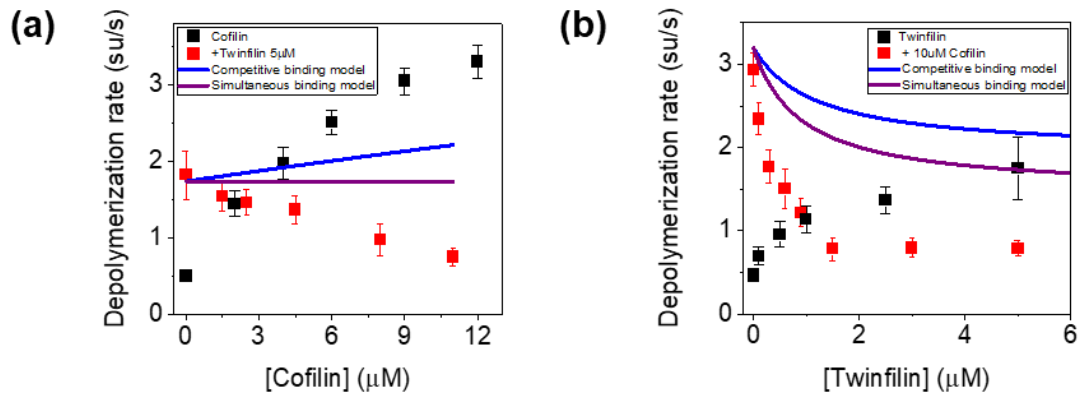

(a) Falsifying competitive and simultaneous binding model for varying concentration of cofilin and fixed twinfilin concentration. Rates (mean ± SD) of barbed end depolymerization in presence of cofilin alone (black symbols) or additionally supplemented with 5 µM twinfilin (red symbols). Number of filaments analyzed for each concentration of cofilin (black symbols, left to right): 40, 60, 60, 60, 60, 60, and 60. Number of filaments analyzed for 5 µM twinfilin and cofilin (red symbols, left to right): 50, 51, 37, 63, 42 and 38. The experimental data was compared to predictions from competitive (blue) and simultaneous (purple) binding model (see SI Equation 10 and 16). (b) Falsifying competitive and simultaneous binding model for varying concentration of twinfilin and fixed cofilin concentration. Rates (mean ± SD) of barbed end depolymerization in presence of twinfilin alone (black symbols) or additionally supplemented with 10 µM cofilin (red symbols). Number of filaments analyzed for each concentration of twinfilin (black symbols, left to right): 62, 64, 62, 59, 67, and 62. Number of filaments analyzed for twinfilin and 10 µM cofilin (red symbols, left to right): 61, 62, 52, 63, 60, 45, 50 and 59. The experimental data was compared to predictions from competitive (blue) and simultaneous binding model (see SI, Equation 10 and 16).

**Detailed of description of models for interaction of one protein, two protein or three proteins with barbed ends.**

**1. Two-state model of depolymerization characterizes single-protein binding to barbed end**

We developed a simple thermodynamic model that employs tools from equilibrium statistical mechanics to characterize how a protein affects the depolymerization of actin barbed end. In this model, a protein $P$ binds to the barbed end of a filament, thereby modulating its observed depolymerization rate. The filament barbed end can either be in a bare state or a protein-bound state, as shown in Supplementary Figure 1a.

The equilibrium probabilities of occupancy of these two states can be computed using equilibrium statistical mechanics (43). The statistical weights corresponding to these discrete states are given by $w_0 = 1$, and $w_1 = C_P/K_{D,P}$. Correspondingly, the partition function of the system is defined as the sum of the statistical weights ($w_i$) i.e.,

$$Z = \sum_i w_i,$$

$$= 1 + \frac{C_P}{K_{D,P}}. \qquad (1)$$

where $C_P$ is the concentration of the protein and dissociation constant is $K_{D,P}$.

Notably, knowledge of partition function fully specifies all the equilibrium statistical properties of the system. Using Equation (1), we obtained the equilibrium probabilities of occupying the bare state ($p_0$) and protein-bound state ($p_1$) which are given by

$$p_0 = \frac{w_0}{Z},$$

$$= \frac{1}{1 + \frac{C_P}{K_D}},$$

$$p_0 = \frac{K_D}{C_P + K_D}. \qquad (2)$$

$$p_1 = \frac{w_1}{Z},$$

$$= \frac{\frac{C_P}{K_D}}{1+\frac{C_P}{K_D}},$$

$$p_1 = \frac{C_P}{C_P + K_D}. \quad (3)$$

The average depolymerization rate $D_P$ of a filament, as measured in experiments, can then be written as:

$$D_P = d_0 p_0 + d_1 p_1, \quad (4)$$

Using equation (4) and substituting for $p_0$ and $p_1$ from Equation (2) and (3), we obtain the average depolymerization rate given by:

$$D_P = d_0 \left(\frac{K_D}{C_P + K_D}\right) + d_1 \left(\frac{C_P}{K_D + C_P}\right) = \frac{d_0 K_D + d_1 C_P}{C_P + K_D}. \quad (5)$$

To quantitively characterize how a single protein affects barbed end depolymerization, we compared the two-state model, as shown in Fig. 1f-h to experimental data. From our experiments we extracted the average depolymerization rate as a function of protein concentration for each protein of interest. We fitted the two-state model to the depolymerization data using Equation (5) to extract the relevant parameters $K_D$, and $d_1$ (see Table.1 in main text). Parameter $d_0 = 0.5$ sub/s was extracted from experimental data.

## 2. Competititve and simultaneous binding models characterize two-protein binding to barbed end

To decipher the mechanism of how two proteins together bind actin barbed end to affect its depolymerization, we developed two broad classes of thermodynamic models, namely competitive and simultaneous binding models. In the competitive model, two proteins bind the barbed end in a mutually exclusive manner, whereby either one of them occupies the barbed end at a given moment. In contrast, in the simultaneous binding model, two proteins can simultaneously occupy the same barbed end. Below, we explain the models in detail.

### 2.1 Competitive binding model

In this model, two proteins A and B bind the barbed end in a competitive manner. As a result, the barbed end can be either in a protein unbound state with probability $p_0$, protein A bound state with probability $p_{1,A}$, or protein B bound state with probability $p_{1,B}$ as shown in figure 2a. Each of these three states is characterized by its distinct depolymerization rate – $d_0$ (free), $d_{1,A}$ (protein A bound) and $d_{1,B}$ (protein B bound).

Here, the equilibrium probabilities $p_0$, $p_{1,A}$, and $p_{1,B}$ can be computed using the partition function as described in the previous section which are given by:

$$p_0 = \frac{1}{\frac{C_B}{K_{D,B}} + 1 + \frac{C_A}{K_{D,A}}},$$

$$p_0 = \frac{K_{D,A} K_{D,B}}{C_B K_{D,A} + K_{D,A} K_{D,B} + C_A K_{D,B}}, \quad (6)$$

$$p_{1,A} = \frac{K_{D,B} C_A}{C_B K_{D,A} + K_{D,A} K_{D,B} + C_A K_{D,B}}, \quad (7)$$

$$p_{1,B} = \frac{K_{D,A} C_B}{C_B K_{D,A} + K_{D,A} K_{D,B} + C_A K_{D,B}}. \quad (8)$$

where $C_A$ and $C_B$ are concentrations of the two proteins, and $K_{D,A/B}$ is the dissociation constant of protein A or B at the barbed end.

Average depolymerization rate when two proteins A and B compete for barbed end of actin filament can then be written as:

$$D_{AB} = d_0 p_0 + d_{1,A} p_{1,A} + d_{1,B} p_{1,B}. \quad (9)$$

Using Eqn. (9) and substituting for $p_0$, $p_{1,A}$, and $p_{1,B}$ from Equation (6), (7), and (8), we obtain the average depolymerization rate when two proteins A and B compete for barbed end of actin filament, given by:

$$D_{AB} = \frac{d_0 K_{D,A} K_{D,B} + d_{1,A} C_A K_{D,B} + d_{1,B} C_B K_{D,A}}{K_{D,A} K_{D,B} + C_A K_{D,B} + C_B K_{D,A}}. \quad (10)$$

The competitive binding model has five parameters namely, $d_0$, $K_{D,A}$, $K_{D,B}$, $d_{1,A}$, and $d_{1,B}$. These parameters are known from our single-protein studies, as elucidated in section.1 (see Table.1 in main text). Parameter $d_0 = 0.5$ sub/s was extracted from experimental data. Using Equation (10), we predicted average depolymerization as a function of concentration of A and B, which was tested against experimental data as shown in Fig. 3b,c, and 4b,c

## 2.2 Simultaneous binding model

In contrast to the competitive binding model, the simultaneous binding model can lead to an additional barbed end state when the two proteins A and B are both simultaneously bound to the barbed end. Consequently, the barbed end can now exist in either in a protein unbound state with probability $p_0$, protein A bound state with probability $p_{1,A}$, protein B bound state with probability $p_{1,B}$, or both protein A and B bound state with probability $p_{2,AB}$ as shown in Supplementary Figure 2b. The rate of depolymerization from these different states are given by – $d_0$ (free), $d_{1,A}$ (protein A bound), $d_{1,B}$ (protein B bound) and $d_{2,AB}$ (both protein A and B simultaneously bound).

Here, the equilibrium probabilities $p_0$, $p_{1,A}$, $p_{1,B}$ and $p_{2,AB}$ can be computed using the partition function as described in the previous section which are given by:

$$p_0 = \frac{K_{D,A}K_{D,B}}{K_{D,A}K_{D,B} + C_A K_{D,B} + C_B K_{D,A} + C_A C_B}, \quad (11)$$

$$p_{1,A} = \frac{C_A K_{D,B}}{K_{D,A}K_{D,B} + C_A K_{D,B} + C_B K_{D,A} + C_A C_B}, \quad (12)$$

$$p_{1,B} = \frac{C_B K_{D,A}}{K_{D,A}K_{D,B} + C_A K_{D,B} + C_B K_{D,A} + C_A C_B}, \quad (13)$$

$$p_{2,AB} = \frac{C_A C_B}{K_{D,A}K_{D,B} + C_A K_{D,B} + C_B K_{D,A} + C_A C_B}. \quad (14)$$

where $C_A$ and $C_B$ are concentrations of the two proteins, and $K_{D,A/B}$ is the dissociation constant of protein A or B at the barbed end.

The average depolymerization for this model is given by:

$$D_{AB} = d_0 p_0 + d_{1,A} p_{1,A} + d_{1,B} p_{1,B} + d_{2,AB} p_{2,AB}. \qquad (15)$$

Using equation (15) and substituting for $p_0$, $p_{1,A}$, $p_{1,B}$, and $p_{2,AB}$ from equation (11), (12), (13), and (14) we obtain the average depolymerization rate, when two proteins A and B can simultaneously occupy barbed end of actin filament, given by:

$$D_{AB} = \frac{d_0 K_{D,A} K_{D,B} + d_{1,A} C_A K_{D,B} + d_{1,B} C_B K_{D,A} + d_{2,AB} C_A C_B}{K_{D,A} K_{D,B} + C_A K_{D,B} + C_B K_{D,A} + C_A C_B}. \qquad (16)$$

We used Equation (16) to predict average depolymerization rate when two proteins bind barbed end cooperatively which is tested against experimental data as shown in figure 4 (b,c). These parameters ($d_{1,A/B}$ and $K_{D,A/B}$) are known from our single-protein studies, as elucidated in section.1 (see Table.1 in main text). Parameter $d_0$ = 0.5 sub/s was extracted from experimental data. Parameter $d_{2,AB}$ was varied, and we found that $d_{2,AB}$ = 11sub/s explained the experimental results best in case of profilin and cofilin.

## 2.3 Revised competitive binding model for Cofilin and Twinfilin

In this model, in addition to the competitive binding of cofilin and twinfilin to the barbed end, cofilin can also bind the sides of the filament. Such side-binding of cofilin can significantly reduce the depolymerization of barbed end, as has been seen previously (28). Here we consider the rate of depolymerization to be zero when two cofilin molecules are bound to the side of the actin filament near the barbed end (one molecule per actin protofilaments) and twinfilin is simultaneously present at the barbed end. In this model, there are nine possible configurations, as shown in Supplementary Figure 1b. There are three states each for i) barbed end binding, ii) barbed end binding and one cofilin molecule binding to one of the protofilaments near the barbed end, and iii) barbed end binding and two cofilin molecules binding to two of the protofilaments near the barbed end. The equilibrium weights of these configurations can be computed using the partition function as described in the previous section which are given by:

$$w_0 = 1,$$

$$w_{1,C} = \frac{C_C}{k_{D,C}},$$

$$w_{1,T} = \frac{C_T}{k_{D,T}},$$

$$w_{S1,0} = \frac{2C_C}{k_{D,SC}},$$

$$w_{S1,C} = \frac{2C_C^2}{k_{D,C}k_{D,SC}},$$

$$w_{S1,T} = \frac{2C_C C_T}{k_{D,T}k_{D,SC}},$$

$$w_{S2,0} = \frac{C_C^2}{k_{D,SC}^2},$$

$$w_{S2,C} = \frac{C_C^3}{k_{D,C}k_{D,SC}^2},$$

$$w_{S2,T} = \frac{\omega C_C^2 C_T}{k_{D,T}k_{D,SC}^2}, \qquad (17)$$

where $C_C$, and $C_T$, are concentrations of cofilin and twinfilin, $K_{D,C/T}$ is the dissociation constant of protein cofilin and twinfilin at the barbed end, $K_{D,SC}$ is the dissociation constant of cofilin at the side of the filament next to the barbed end, and ω is the cooperativity between side binding of two cofilin molecules to the two of the protofilaments near the barbed end and barbed end-bound twinfilin.

The rate of depolymerization from these different states are given by – $d_0$ (free), $d_{1,C}$ (cofilin bound), $d_{2,T}$ (twinfilin bound) and $d_{2,TC}$ (side binding of cofilin to the two of the protofilaments near the barbed end and barbed end bound twinfilin).

The average depolymerization for this model can be calculated using corresponding weights and depolymerization rates of the nine states as given by:

$$D_{AB} = \frac{d_0 w_0 + d_{1,C} w_{1,C} + d_{1,T} w_{1,T} + d_0 w_{S1,0} + d_{1,C} w_{S1,C} + d_{1,T} w_{S1,T} + d_0 w_{S2,0} + d_{1,C} w_{S2,C} + d_{2,TC} w_{S2,T}}{\sum_i w_i}.$$

(18)

We used Equation (18) after plugging in weights from Equation (17) to predict average depolymerization rate when cofilin and twinfilin interact with barbed end which is tested against experimental data as shown in figure 5 (b,c). These parameters ($d_{1,C/T}$ and $K_{D,C/T}$) are known from our single-protein studies, as elucidated in section.1 (see Table.1 in main text). Parameter $d_0 = 0.5$ sub/s was extracted from experimental data. The parameter $K_{D,SC} = 181 \mu M$ was taken from a previous study (49), and $\omega$ was varied and we found that $\omega = 1084.1$ explained the experimental results best. The depolymerization rate when two cofilin molecules bind two of the protofilaments near the barbed end and simultaneously twinfilin is bound to the barbed end is taken to be zero, i.e., $d_{2,TC}=0$.

### 3. Three protein case

We combined the pair-wise interactions of profilin, twinfilin, and cofilin to predict the average depolymerization rate in the presence of three proteins, as shown in Supplementary Figure 1c. As before, the equilibrium weights of the various states are computed from the partition function. These weights are given by: $w_0 = 1$,

$$w_{1,C} = \frac{C_C}{k_{D,C}},$$

$$w_{1,P} = \frac{C_P}{k_{D,P}},$$

$$w_{1,T} = \frac{C_T}{k_{D,T}},$$

$$w_{2,PC} = \frac{C_C C_P}{k_{D,P} k_{D,C}},$$

$$w_{S1,0} = \frac{2 C_C}{k_{D,SC}},$$

$$w_{S1,C} = \frac{2 C_C^2}{k_{D,C} k_{D,SC}},$$

$$w_{S1,T} = \frac{2C_C C_T}{k_{D,T} k_{D,SC}},$$

$$w_{S2,0} = \frac{C_C^2}{k_{D,SC}^2},$$

$$w_{S2,C} = \frac{C_C^3}{k_{D,C} k_{D,SC}^2},$$

$$w_{S2,T} = \frac{\omega C_C^2 C_T}{k_{D,T} k_{D,SC}^2}, \qquad (19)$$

The rate of depolymerization from these different states are given by $d_0$ (free), $d_{1,C}$ (cofilin bound), $d_{1,P}$ (profilin bound), $d_{1,T}$ (twinfilin bound), $d_{2,PC}$ (both profilin and cofilin simultaneously bound), and $d_{2,TC}$ (twinfilin is bound on the barbed end and cofilin is occupying two binding sites on the side) depolymerization rates. Here $C_P$, $C_C$, and $C_T$ are concentrations of the three proteins, $K_{D,P/C/T}$ is the dissociation constant of profilin, cofilin, and twinfilin at the barbed end, and $K_{D,SC}$ is the dissociation constant of cofilin on the side of the filament next to the barbed end.

The average depolymerization for this model is given by:

$$D_{CPT} = \frac{d_0 w_0 + d_{1,C} w_{1,C} + d_{1,P} w_{1,P} + d_{1,T} w_{1,T} + d_{2,PC} w_{2,PC} + d_0 w_{S1,0} + d_{1,C} w_{S1,C} + d_{1,T} w_{S1,T} + d_0 w_{S2,0} + d_{1,C} w_{S2,C} + d_{2,TC} w_{S2,T}}{\sum_i w_i}.$$

(20)

We used Equation (20) after plugging in the weights from Equation (19) to predict the average depolymerization rate in the presence of three proteins which was tested against experimental data as shown in figure 6 (b,c). These parameters ($d_{1,C/P/T}$ and $K_{D,P/C/T}$) are known from our single-protein studies, as elucidated in section.1 (see Table.1 in main text). Parameter $d_0$ = 0.5 sub/s was extracted from experimental data. Parameters $d_{2,PC}$ = 11 sub/s, $d_{2,TC}$ = 0 sub/s, $K_{D,SC}$ = 181 µM, and $\omega$ = 1084.1 were obtained from the previous sections.